# Evaluating m-learning in Saudi Arabian higher education: a case study

**Salem Alkhalaf**

Computer Department, College of Science and Arts, Qassim University
Alrass City, Qassim, KSA

## Abstract

***Nowadays, mobile devices have become increasingly a part of education for those who study or teach at the university level and school levels. The support of electronic learning (e-learning) is essential to making mobile learning (m-learning) successful. This paper presents a study that applies m-learning to a course at Qassim University, where 100 students attended during the academic year 2014, including summer courses. The study aims to demonstrate that m-learning provides students with the ability to engage in reflective thinking, to share information among peers and to facilitate the construction of social knowledge. 90 student questionnaires were filled correctly, remaining 10 had various anomalies and thus were not considered. The study also aims to demonstrate that learning provides methods for education strategies to be easily and rapidly applied. Such strategies include team work, time management, etc. In order to judge the feasibility of applying m-learning widely, a questionnaire was developed. The results indicate that m-learning helps make the process of education more convenient than in the past. A major disadvantage of applying m-learning is inadequate wireless network bandwidth.***

**Keywords:** m-learning, e-learning, learning experiences.

## 1. Introductio

Problem-based learning (PBL) is an education approach that has developed around the exploration of problems along with their resolution. Rapid development and availability of electronic and mobile devices with superior capabilities has stimulated learners to apply PBL in their lives, inside and outside of schools [7]. The use of devices with web accessibility is increasing rapidly and affecting all levels of education. Availability of such devices is beginning to change the way people undertake the process of learning [33]. Learning and/or teaching processes can now extend outside labs and classrooms through cellular devices that include smart phones, notebooks, laptops, and personal digital assistants (PDAs) [34]. Learners can easily obtain notes regarding any discussions or lectures and assignments with such cellular devices [33].

Mobile learning (m-learning) approach has been celebrated as having the potential to become an important part of the process of attaining education and, ultimately, the future of the learning domain [1]. Cell phone applications for the process of learning have become even more popular, along with discussions on various mobile learning plans [2], [3], [4], [5], [6]. Construction of social knowledge behaviour patterns with regard to the environment of electronic learning (e-learning) and m-learning has been discussed and investigated [7]. It is well known that when students use m-learning, they are provided with the ability to engage in reflective thoughts, to share information among peers and to facilitate the construction of social knowledge [7], [8]. The effect of technology with regard to mobile learning from the perspective of the educator, as well as the learner has been analysed and investigated. The study by Abachi and Muhammad [9] considered the undergraduate and graduate students. The effectiveness of m-learning sessions has also been questioned, particularly with regard to m-learning's reliability, as well as how valid and accurate the content deliverables are. Emma and Walid [10] studied these factors and investigated their effects on m-learning techniques. The application of m-learning within the classroom setting and the analysis of the educational experiences associated with it have also been discussed [11]. Augmented reality, as well as m-learning has been implemented to determine students' performances in higher education.

Due to rapid adoption of m-learning throughout the world, research focusing on factors influencing the adoption of m-learning have been undertaken [12], [13]. An improvement of student performance with the use of m-learning through auto-assessment in secondary, as well as tertiary education has also been discussed [13]. M-learning is assessed based on students' acceptance of the process and the benefits derived from it.

The opinions of teacher candidates regarding applying mobile devices in the learning process have also been studied [14]. As students use the Internet extensively by laptop or mobile devices, application of m-learning has become important and attractive to many teachers. The differences among applying m- and e-learning have been discussed and investigated [15]. A major advantage of mobile devices is the fact that they are not very bulky to carry around, yet include numerous features, and this increases public interest. This also encourages people to apply them in the learning process.

The design principles of m-learning for English as a second language (ESL) studies have been investigated and discussed [16]. The effectiveness of m-learning through podcast lectures for the purpose of revision has also been analysed [17].

The appropriate technological framework for m- and e-learning can be made available through public key infrastructure (PKI) and attribute certificates (ACs). These effectively support authentication, as well as authorization services and offer a sense of trust to the service providers and learners. The potential function of ACs to sustain m-learning and e-learning has been investigated and discussed [18]. Case studies on the applications of m-learning have received attention from Dearnley, Haigh, and Fairhall [19]; Binsaleh and Binsaleh [20]; Jones, Scanlon, and Clough [21]; and Pimmer et al. [22].

A case study for applying cellular technologies to the evaluation, as well as the process of learning within practice settings was conducted by Dearnley, Haigh, and Fairhall [19]. The study focused on cellular technology application in terms of the evaluation of students within health and social care settings. M-learning strategies have also been used to overcome societal problems and limitations, such as those caused by unrest in the four southernmost provinces of Thailand (Binsaleh & Binsaleh, 2013). This paper identifies the factors affecting improvements in and overcoming educational problems via m-learning. Mobile technologies can provide high-level educational processes in a variety of circumstances due to their portability, enabling them to be used by learners regardless of location [21]. Pimmer et al. [22] investigated the application of m-learning in nursing education and practices in remote and rural areas in South Africa. Pimmer et al. concluded that mobile phones are highly effective in varied and changeable learning environments. The appropriation of mobile phone utilization has been evaluated and investigated [23]. The associated applications and usage have been investigated. Rapid development of mobile technology pervasive learning can enhance the effectiveness and accessibility of learning activities in the future. An assessment of student's perception of implementing augmented reality in learning processes has been discussed and investigated [24]. Teaching and learning this topic has been investigated using a developed mobile phone. The ADDIE model, which consists of five phases: Analysis, Design, Development, Implementation and Evaluation has been utilized to conduct the study. The study found that students' perception of the use of augmented reality is positive based on the average mean score. The educational mobile application has been used to develop the activity-based learning model for improvement discipline of elementary school students. Reviews of literature and experts' interview have been used to develop the model [25]. The model evaluated by 30 elementary school students, followed by approval from the experts. The obtained results show that this model consists of five components along with four steps. Mobile devices with highest technologies have been extended to cover many areas, such as in language learning and education [26]. The IBan language has been chosen to be implemented as a multimedia-based mobile application, and obtained results have shown that this application is a promising development for future research [27]. Mobile learning is applied based on the use of a set of specific mobile applications, which possesses similar features and specific characteristics according to learning goals. The software Product Line (SPL) allows mass customization and systematic derivation of products, such as applying mobile devices in the learning process. Development of the MSP Learning process has been discussed on the implementation phase. The facilitation of mobile learning processes in technical vocational and engineering education have been discussed and investigated [28]. This technique includes the distribution of classical learning materials, such as text, pictures, videos, or simulations. E-Learning is well known and broadly used in different kinds of education, this is not true for mobile learning (m-Learning), even though it offers new possibilities as a supplement beyond micro learning. The importance of applying m-learning comes from the advantages of its availability anytime anywhere learning which is important for students who frequently change their location and need access to learning material as they move from one place to another [29]. Mobile devices are mostly cheap, portable, and flexible, have no start-up time, and require virtually no maintenance. Also, mobile technology seems to be very attractive to students and usable in the learning process.

This paper aims to investigate the feedback from students' on the mobile learning process through applying such techniques in one subject for full academic year, including summer classes. The questionnaire is designed to measure a student's acceptability and convenience with the technique. The reflection was very clear not only from the questionnaire results but also the final results of students were much better than before when applying such classical learning approaches.

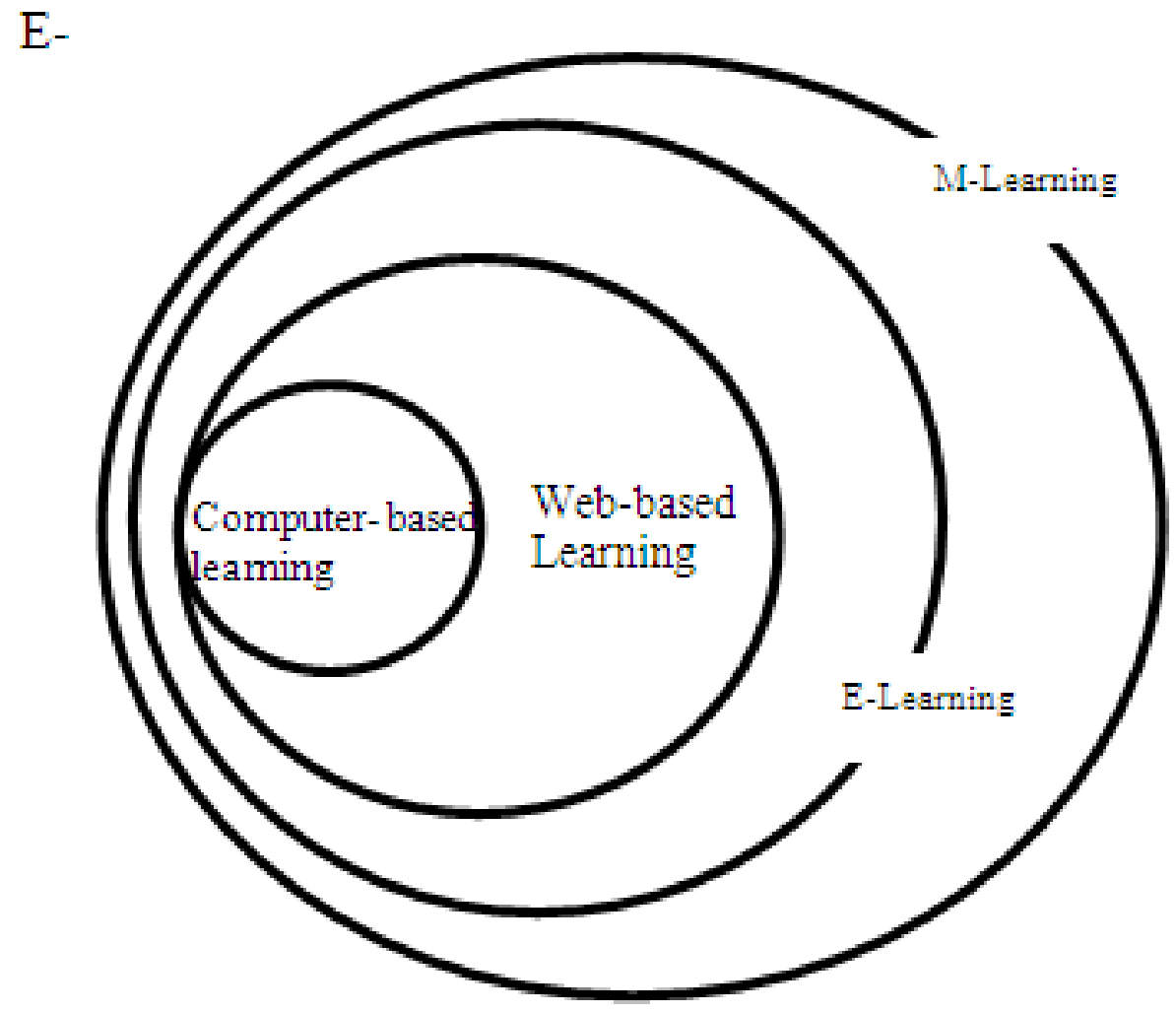

**Figure 1** Relationship of m-learning to e-learning along with other learning concepts.

## 2. DEFINITION OF M-LEARNING

### A. DEFINITION, CONSIDERING THE LEARNER

According to Rüdel [30], the most important aspect with respect to the definition of m-learning is its connection to e-learning along with various learning concepts, as shown in Figure. 1. Rüdel's definition of m-learning (in terms of the learner) hinges on two key factors:

- Anywhere — location independent
- Anytime — time independent

### B. DEFINITION, CONSIDERING THE LEARNING PROCESS

Rüdel's [30] definition of m-learning (in terms of the learning process) also hinges on two key factors:

- situated/authentic
- fieldwork/workplace

### C. DEFINITION, CONSIDERING THE DEVICES AND THEIR PORTABILITY

Rüdel [23] specifies that m-learning can be accomplished via an MP3 player, phone, PDA, laptop, and computer.

## 3.M-LEARNING DEVICES

Present-day mobile devices most often used for m-learning include the following:

1) Cell phones, wireless application protocol (WAP) phones, and 3G phones
2) Tablet PCs
3) PDAs
4) E-book readers
5) Hybrid devices

## 4.M-LEARNING PROS AND CONS

Studies have determined that m-learning technique has both, advantages and disadvantages.

### A. ADVANTAGES OF M-LEARNING

Boyes [31] lists numerous advantages to m-learning. Eleven significant ones are as follows:

1) M-learning can be easily accessed anytime, anywhere.
2) M-learning very easily enables training.
3) M-learning is easy to control.
4) M-learning can be used in the "dead times."
5) M-learning provides many different learning styles.
6) M-learning improves social learning.
7) M-learning encourages reflection.
8) M-learning supports decision making.
9) M-learning improves learner confidence.
10) M-learning helps learners with planning.
11) M-learning directs interaction with learners.

### B. DISADVANTAGES OF M-LEARNING

Four main disadvantages to m-learning are as follows:

1) High cost
2) Size of device
3) Battery life
4) Usability

## 5.AIMS AND OBJECTIVES

The researcher was motivated to conduct the study when preparing for Qassim University's Communication Skills course. The researcher was preparing course documents as PDF files and establishing a group to provide online discussions and explain potentially confusing points.

**The study aims to undertake the following:**

- Examine the feasibility of applying m-learning.
- Indicate students' perceptions toward application of e-learning.
- Determine how the application of m-learning can provide more learning activities for students.
- Determine how m-learning techniques can be applied to a wide range of subjects.
- Determine whether instructors' views on m-learning or e-learning can affect student perceptions.

## 6.RESEARCH QUESTIONS

During the summer semester in 2014 at Qassim University, fifteen students in the Communication Skills course were requested to fill in a questionnaire to assess their experiences of m-learning. The following questions were considered:

1) What mobile devices do students use, and what are the students' user patterns?
2) What are the students' reflections on the application of m-learning?
3) Does the application of m-learning assist with student motivation?
4) Does the application of m-learning modify student skills?
5) Is m-learning more flexible, easier and more successful than other techniques?

## 7.Results and discussion

The questionnaire covered many aspects of the feasibility of applying m-learning. The results of the first part in the questionnaire proved that most of the students are using mobile phones. Also, they are considering it as a suitable tool for learning. Most of them are using the device 4 times or more daily and most of them spent about 3-hrs per day using the device for learning purposes.

The SPSS statistical program has been used to analyse the data from part 2 in the questionnaire through part 5 which includes 20 questions. Figs 2 shows student reflections on the applications of m-learning. Fig. 3 depicts student reflections on m-learning and motivations. Fig. 4 shows student reflections on m-learning and skills. Fig. 5 shows student reflections on m-learning and successful use of the technology. Table 1 through 4 shows the results of the questionnaire. Table 5 through 8 provides the Likert scale analysis for the collected data to indicate the feasibility of the study.

**Table1.** Student reflections on the application of m-learning.

| | | Totally agree | Agree | Neutral | Disagree | Totally disagree |
|---|---|---|---|---|---|---|
| 1 | The m-learning technique | 64 | 14 | 12 | 0 | 0 |
| | attracts me. | | | | | |
| 2 | M-learning is an easy technique. | 78 | 6 | 6 | 0 | 0 |
| 3 | M-learning assists me. | 76 | 12 | 2 | 0 | 0 |
| 4 | M-learning is available anytime. | 72 | 8 | 12 | 0 | 0 |
| 5 | M-learning is my first choice. | 73 | 13 | 4 | 0 | 0 |

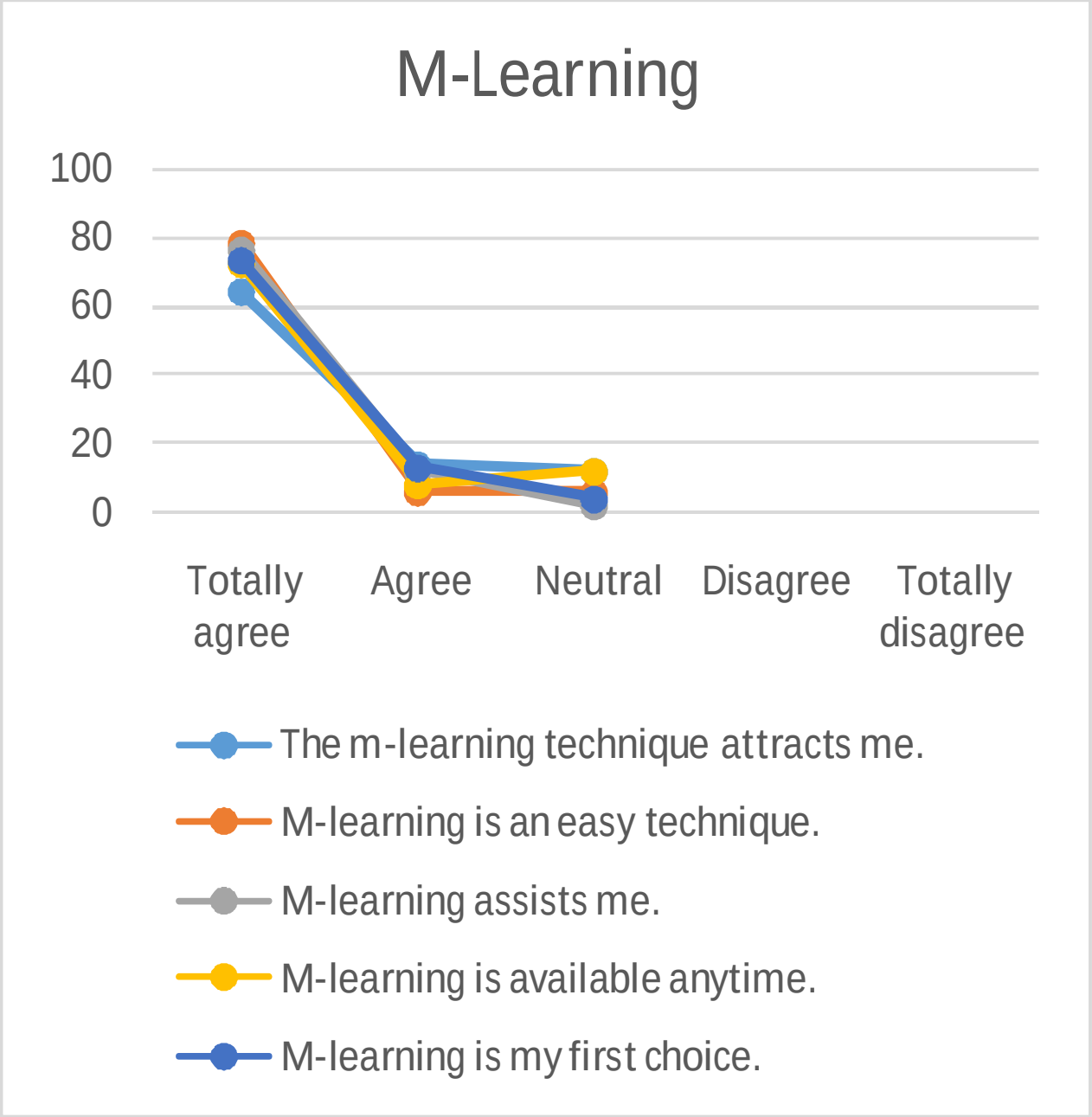

**FIGURE. 2** STUDENT REFLECTIONS ON THE APPLICATIONS OF M-LEARNING

**Table 2** Student reflections on m-learning and motivation.

| | | Totally agree | Agree | Neutral | Disagree | Totally disagree |
|---|---|---|---|---|---|---|
| 1 | M-learning is an exciting technique. | 71 | 13 | 6 | 0 | 0 |
| 2 | The m-learning technique increased my learning aptitude. | 75 | 13 | 2 | 0 | 0 |
| 3 | M-learning is an easier technique than other learning techniques. | 79 | 6 | 5 | 0 | 0 |
| 4 | M-learning saves me time. | 77 | 10 | 3 | 0 | 0 |
| 5 | M-Learning can be accessed easily. | 74 | 13 | 3 | 0 | 0 |

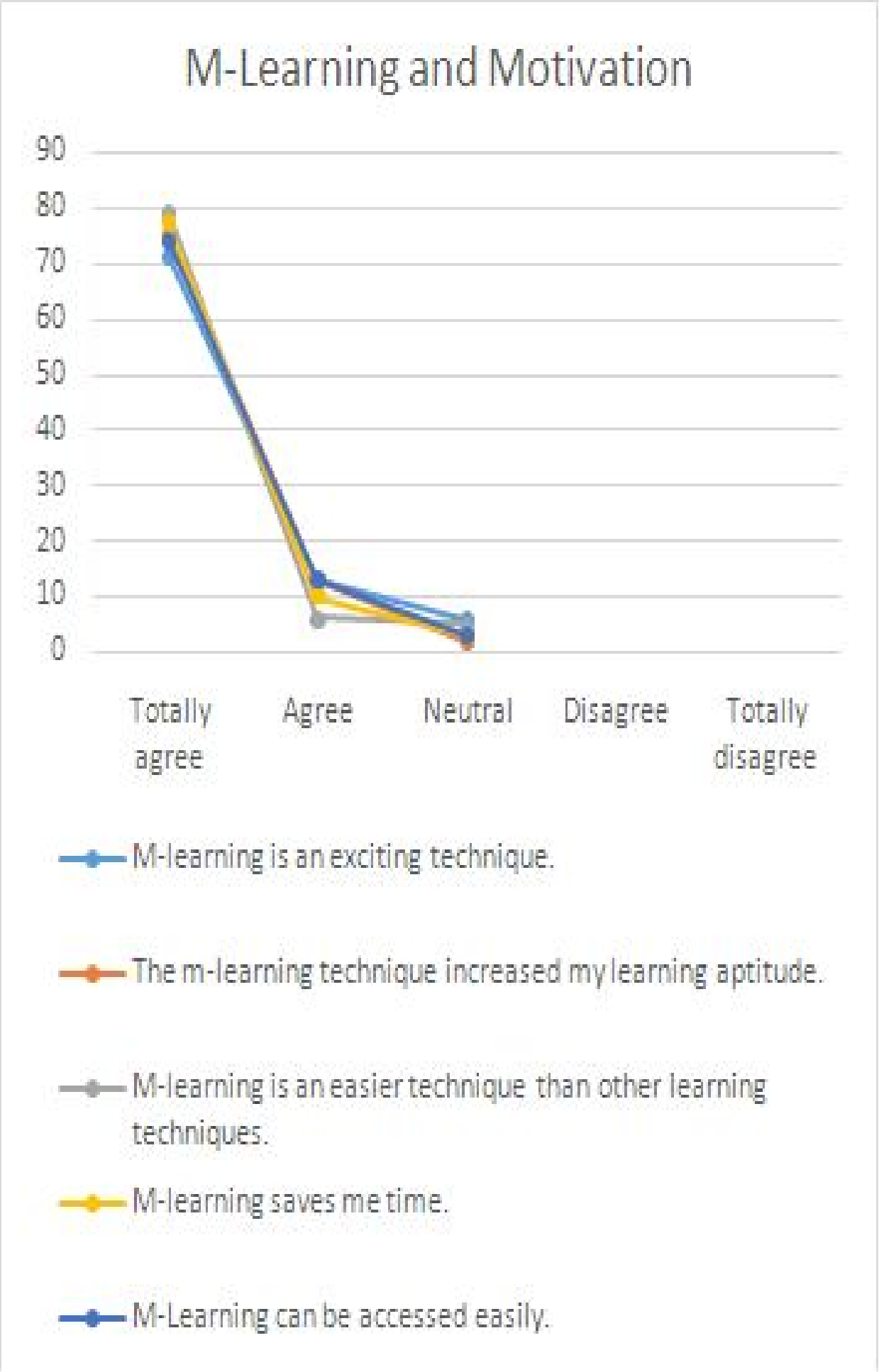

**FIGURE. 3** STUDENT REFLECTION ON M-LEARNING AND MOTIVATIONS

**Table 3.** Student reflections on m-learning and skills.

| No | | Totally agree | Agree | Neutral | Disagree | Totally disagree |
|---|---|---|---|---|---|---|
| 1 | M-learning helps me greatly. | 74 | 12 | 4 | 0 | 0 |
| 2 | M-learning improved my skills in mobile applications. | 73 | 12 | 5 | 0 | 0 |
| 3 | M-learning supports my ability to use mobile devices easily. | 79 | 7 | 4 | 0 | 0 |
| 4 | M-learning directs my skills in a beneficial way. | 80 | 7 | 3 | 0 | 0 |
| 5 | M-learning is an exciting way to update my knowledge about mobile applications. | 82 | 2 | 6 | 0 | 0 |

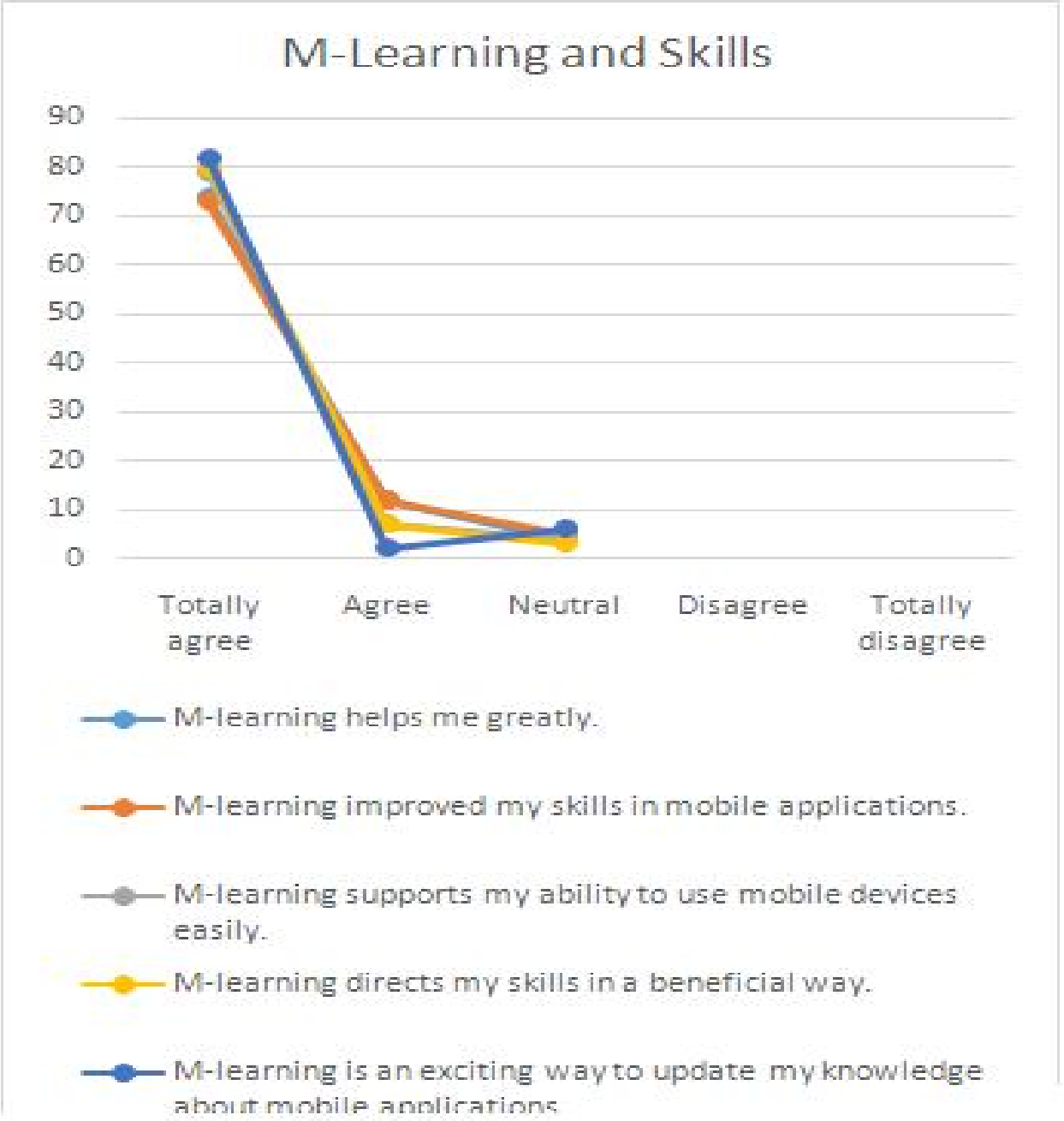

**FIGURE. 4** STUDENT REFLECTIONS ON M-LEARNING AND SKILLS

**Table 4.** Student reflections on m-learning and successful use.

| NO | | Totally agree | Agree | Neutral | Disagree | Totally disagree |
|---|---|---|---|---|---|---|
| 1 | M-learning is a successful technique. | 77 | 12 | 1 | 0 | 0 |
| 2 | M-learning much easier than other techniques. | 83 | 3 | 4 | 0 | 0 |
| 3 | M-learning has many advantages. | 69 | 14 | 7 | 0 | 0 |
| 4 | M-learning is easy to apply to a wide range of subject matter. | 74 | 14 | 2 | 0 | 0 |
| 5 | M-learning is easy to understand. | 75 | 11 | 4 | 0 | 0 |

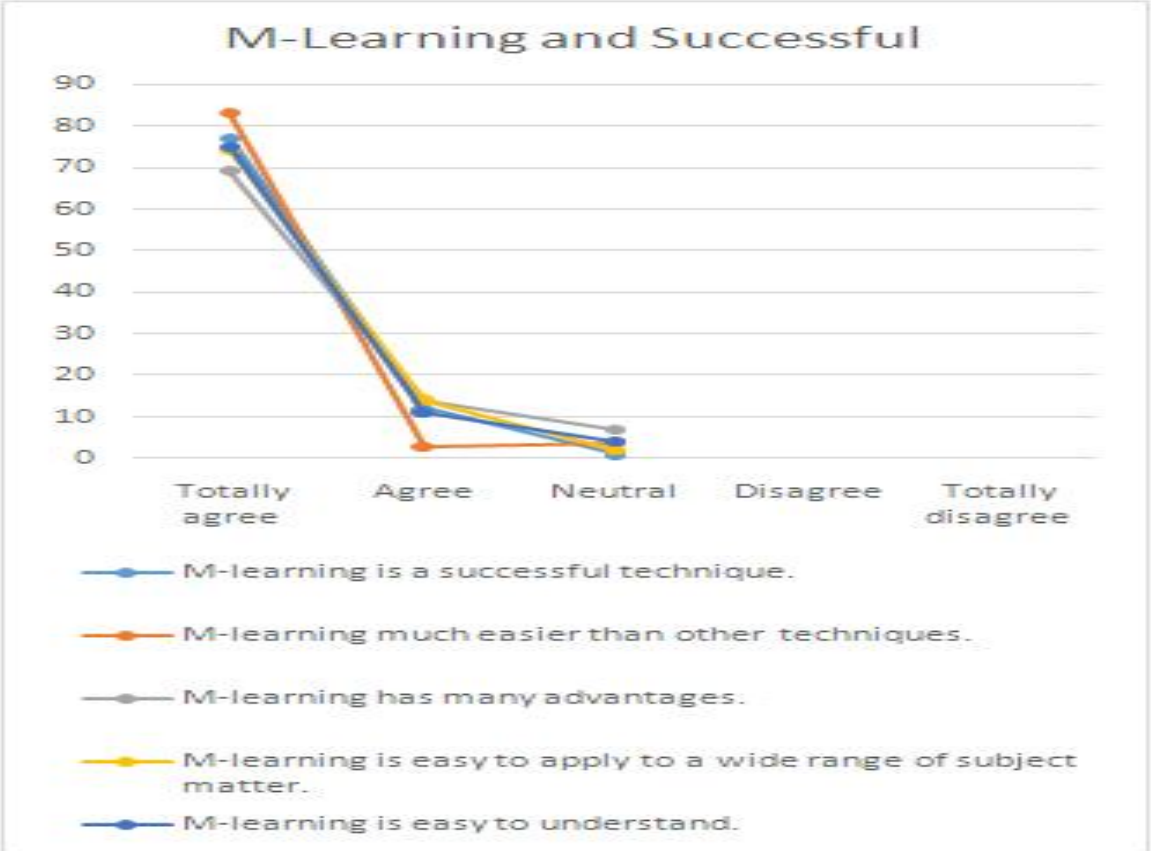

**Figure. 5** Student reflections on m-learning and successful use

**Table 5 Likert scale for Student reflections on the application of m-learning**

| weighted mean | Level |
|---|---|
| from 1.00 to 1.79 | **Completely disagree 1** |
| from 1.80 to 2.59 | **Disagree 2** |
| from 2.60 to 3.39 | **Neutral 3** |
| from 3.40 to 4.19 | **Agree 4** |
| from 4.25 to 5.00 | **Completely agree 5** |

**Table 6** Likert scale for the student reflections on m-learning and motivation

| Measurement | Totally disagree | **Disagree** | **Neutral** | **Agree** | **Totally agree** | Mean | St. Dev. | Result |
|---|---|---|---|---|---|---|---|---|
| Freq. | 0 | 0 | 12 | 14 | 64 | 4.5778 | .71858 | **Completely agree** |
| Prop. | 0 | 0 | 13.3 | 15.6 | 71.1 | | | |
| Freq. | 0 | 0 | 6 | 6 | 78 | 4.8000 | .54464 | **Completely agree** |
| Prop. | 0 | 0 | 6.7 | 6.7 | 86.7 | | | |
| Freq. | 0 | 0 | 2 | 12 | 76 | 4.8222 | .43904 | **Completely agree** |
| Prop. | 0 | 0 | 2.2 | 13.3 | 84.4 | | | |
| Freq. | 0 | 0 | 12 | 8 | 72 | 4.6889 | .66442 | **Completely agree** |
| Prop. | 0 | 0 | 13.3 | 8.9 | 80 | | | |
| Freq. | 0 | 0 | 4 | 13 | 73 | 4.7667 | .52037 | Completely agree |
| Prop. | 0 | 0 | 4.4 | 14.4 | 81.1 | | | |
| Freq. | 0 | 0 | 7.5 | 10.6 | 72.6 | 4.7311 | .5774 | Completely agree |
| Prop. | 0 | 0 | 8.3 | 11.8 | 80.7 | | | |

Application of m-learning:
1) The m-learning technique attracts me. 2) M-learning is an easy technique. 3) M-learning assists me. 4) M-learning is available anytime. 5) M-learning is my first choice. 6) Attitude Results

**Table 7** Likert scale for the student reflections on m-learning and skills

| No | Measurement | Totally disagree | **Disagree** | **Neutral** | **Agree** | **Totally agree** | Mean | St. Dev. | Result |
|---|---|---|---|---|---|---|---|---|---|
| 1 | Freq. | 0 | 0 | 6 | 13 | 71 | 4.7444 | .53129 | Completely agree |
| 2 | Prop. | 0 | 0 | 6.6 | 14.5 | 78.9 | | | |
| 3 | Freq. | 0 | 0 | 2 | 13 | 75 | 4.8111 | .44707 | Completely agree |
| 4 | Prop. | 0 | 0 | 2.2 | 14.5 | 83.3 | | | |
| 5 | Freq. | 0 | 0 | 5 | 6 | 79 | 4.8222 | .51007 | Completely agree |
| 6 | Prop. | 0 | 0 | 5.6 | 6.6 | 89.8 | | | |
| 7 | Freq. | 0 | 0 | 3 | 10 | 77 | 4.8222 | .46393 | Completely agree |
| 8 | Prop. | 0 | 0 | 3.3 | 11.1 | 85.6 | | | |
| 9 | Freq. | 0 | 0 | 3 | 13 | 74 | 4.7889 | .48562 | Completely agree |
| 10 | Prop. | 0 | 0 | 3.3 | 14.5 | 82.2 | | | |
| 11 | Freq. | 0 | 0 | 3.2 | 11 | 75.2 | 4.7977 | .4876 | Completely agree |
| 12 | Prop. | 0 | 0 | 3.5 | 12.2 | 83.5 | | | |

**Reflections on m-learning and motivation:**
1) M-learning is an exciting technique. 2) The m-learning technique increased my learning aptitude. 3) M-learning is an easier technique than other learning techniques. 4) M-learning saves me time. 5) M-Learning can be accessed easily. 6) M-learning is an exciting technique. 7) The m-learning technique increased my learning aptitude. 8) M-learning is an easier technique than other learning techniques. 9) M-learning saves me time. 10) Attitude Results.

**Table 8** Likert scale for the student reflections on m-learning and successful use

| No | Measurement | Totally disagree | **Disagree** | **Neutral** | **Agree** | **Totally agree** | Mean | St. Dev. | Result |
|---|---|---|---|---|---|---|---|---|---|
| 1 | Freq. | 0 | 0 | 4 | 12 | 74 | 4.7778 | .51446 | Completely agree |
| | Prop. | 0 | 0 | 4.44 | 13.3 | 82.2 | | | |

| No | Measurement | Totally disagree | Disagree | Neutral | Agree | Totally agree | Mean | St. Dev. | Result |
|---|---|---|---|---|---|---|---|---|---|
| 2 | Prop. | 0 | 0 | 5.5 | 13.3 | 81.1 | 4.7556 | .54692 | Completely agree |
| | Freq. | 0 | 0 | 5 | 12 | 73 | | | |
| 3 | Prop. | 0 | 0 | 4.5 | 7.8 | 87.7 | 4.8333 | .47993 | Completely agree |
| | Freq. | 0 | 0 | 4 | 7 | 79 | | | |
| 4 | Prop. | 0 | 0 | 3.3 | 7.8 | 88.9 | 4.8556 | .43862 | Completely agree |
| | Freq. | 0 | 0 | 3 | 7 | 80 | | | |
| 5 | Prop. | 0 | 0 | 6.7 | 2.2 | 91.1 | 4.8444 | .51736 | Completely agree |
| | Freq. | 0 | 0 | 6 | 2 | 82 | | | |
| 6 | Prop. | 0 | 0 | 4.9 | 6.6 | 86.2 | 4.81334 | 0.499458 | Completely agree |
| | Freq. | 0 | 0 | 4.4 | 6 | 77.6 | | | |

**Reflections on m-learning and skills**

1) M-learning helps me greatly. 2) M-learning improved my skills in mobile applications. 3) M-learning supports my ability to use mobile devices easily. 4) M-learning directs my skills in a beneficial way. 5) M-learning is an exciting way to update my knowledge about mobile applications. 7) M-learning helps me greatly. 8) M-learning improved my skills in mobile applications. 9) M-learning supports my ability to use mobile devices easily. 10) M-learning directs my skills in a beneficial way. 11) Attitude Results

**Table 9** The Likert scale analysis indicating that m-learning technique is more flexible, easier, and more successful than traditional techniques as it helps students in a different manner.

| No | Measurement | Totally disagree | Disagree | Neutral | Agree | Totally agree | Mean | St. Dev. | Result |
|---|---|---|---|---|---|---|---|---|---|
| 1 | Freq. | 0 | 0 | 1 | 12 | 77 | 4.8444 | .39409 | Completely agree |
| 2 | Prop. | 0 | 0 | 1.1 | 13.3 | 85.5 | | | |
| 3 | Freq. | 0 | 0 | 4 | 3 | 83 | 4.8778 | .44540 | Completely agree |
| 4 | Prop. | 0 | 0 | 4.4 | 3.3 | 92.2 | | | |
| 5 | Freq. | 0 | 0 | 7 | 14 | 69 | 4.6889 | .61158 | Completely agree |
| 6 | Prop. | 0 | 0 | 7.8 | 15.5 | 76.6 | | | |
| 7 | Freq. | 0 | 0 | 2 | 14 | 74 | 4.8000 | .45469 | Completely agree |
| 8 | Prop. | 0 | 0 | 2.2 | 15.5 | 82.2 | | | |
| 9 | Freq. | 0 | 0 | 4 | 11 | 75 | 4.7889 | .50823 | Completely agree |
| 10 | Prop. | 0 | 0 | 4.5 | 12.2 | 83.3 | | | |
| 11 | Freq. | 0 | 0 | 3.8 | 10.8 | 75.6 | 4.8 | .4828 | Completely agree |
| 12 | Prop. | 0 | 0 | 4.2 | 12 | 84 | | | |

**Reflections on m-learning and successful use:**

1. M-learning is a successful technique. 2) M-learning much easier than other techniques. 3) M-learning has many advantages. 4) M-learning is easy to apply to a wide range of subject matter. 5) M-learning is easy to understand. 6) M-learning is a successful technique. 7) M-learning much easier than other techniques. 8) M-learning has many advantages. 9) M-learning is easy to apply to a wide range of subject matter. 10) Attitude Results

## 8.LIMITATIONS OF THE STUDY

The study had a few limitations. Student survey was a relatively small sample size (N=100), so the students' responses may not be indicative of most students. The study is applied over the period of 1-year only, responses and results from the surveyed sample are subject to change. Some of the mobile devices had limitations in applying m-learning techniques. Lastly, one could not assume all students had a natural aptitude for applying mobile devices in learning.

## 9.CONCLUSION

The study proves that students continue their learning activities outside their classes anytime, provided a wi-fi network is available. Student's reflections on applying mobile devices in learning are highly noteworthy. The study also demonstrates that m-learning strategies have become crucial techniques that may be applied to a wide range of studies. Finally, the study clearly indicates that m-learning has many benefits over traditional classroom-centric learning. The Likert scale analysis and SPSS analysis concluded that the m-learning is an essential tool, based to the results of the questionnaire and may be considered a promising technology due to continuous development and evolvement in software programs and mobile platforms.

## References

[1] Ozuorcun, N. C., &Tabak, F.(2012).Is M-learning versus E-learning or are they supporting each other? Procedia—Social and Behavioral Sciences, 46, 299–305.
[2] Chen, H. R., & Huang, H. L. (2010). User Acceptance of mobile knowledge management learning system: design and analysis. Educational Technology & Society, 13(3), 70–77.
[3] Sung, Y. T., Hou, H. T., Liu, C. K., & Chang, K. E. (2010). Mobile guide system using problem-solving strategy for museum learning: a sequential learning behavioral pattern analysis. Journal of Computer Assisted Learning, 26(2), 106–115.
[4] Cheng, S. C., Hwang, W. Y., Wu, S. Y., Shadiev, R., &Xie, C. H. (2010). A mobile device and online system with contextual familiarity and its effects on English learning on campus. Educational Technology & Society, 13(3), 93–109.
[5] Chang, C. S., Chen, T. S., &Hsu,W. H. (2011). The study on integratingWebQuest with mobile learning for environmental education. Computers & Education, 57(1), 1228–1239.
[6] Sandberg, J., Maris, M., & de Geus, K. (2011). Mobile English learning: an evidence-based study with fifth graders. Computers & Education, 57(1), 1334–1347.
[7] Lan,Y.-F., Tsa, P.-W.,Yang, S.-H.,&Hung,C.-H.(2012)Comparing the social knowledge construction behavioral patterns of problem-based online asynchronous discussion in e/m-learning environments.Computers & Education 59, 1122–1135.
[8] Nik Mastura Nik Mohammad, Mohd Nor Mamat,&PosiahMohd Isa.(2012).M-learning in Malaysia: Challenges and Strategies.Procedia—Social and Behavioral Sciences 67, 393–401.
[9] Abachi, H. R., &Muhammad, G. (2014).The impact of m-learning technology on students and educators.Computers in Human Behavior 30, 491–496.
[10] Emma NuraihanMior Ibrahim, &Walid, N. ( 2014).Trust contributing factors in m-learning technology.Procedia—Social and Behavioral Sciences 129, 554–561.
[11] Cadavieco, J. F., de Fatima Goulao, M.,&Costales, A. F.(2012).Using augmented reality and m-learning to optimize students' performance in higher education.Procedia—Social and Behavioral Sciences 46, 2970–2977.
[12] Yong,L., Hongxiu,L., &Christer, C.(2010)Factors driving the adoption of m-learning: an empirical study.Computers& Education 55, 1211–1219.
[13] De-Marcos, L., Hilera, J. R., Barchino, R., Jiménez, L, Martínez, J. J., Gutiérrez, J. A., Gutiérrez, J. M., and Otón, S.(2010).An experiment for improving students' performance in secondary and tertiary education by means of m-learning auto-assessment.Computers& Education 55, 1069–1079.
[14] Ozdamli, F., Soykan, E., &Yıldız, E. P.(2013). Are computer education teacher candidates ready for m-learning? Procedia—Social and Behavioral Sciences 83, 1010–1015.
[15] Korucu, A. T., &Alkan, A.(2011)Differences between m-learning (mobile learning) and e-learning,basic terminology and usage of m-learning in education.Procedia Social and Behavioral Sciences 15, 1925–1930.
[16] Kwona, S., &Lee, J. E.(2010).Design principles of m-learning for ESL.Procedia Social and Behavioral Sciences 2, 1884–1889.
[17] Evans, C. (2008).The effectiveness of m-learning in the form of podcast revision lectures in higher education.Computers& Education 50, 491–498.
[18] Kambourakis, G. Kontoni, D.-P. N., Rouskas, A., &Gritzalis, S. (2007). A PKI approach for deploying modern secure distributed e-learning and m-learning environments.Computers& Education 48, 1–16.
[19] Dearnley, C., Haigh, J., &Fairhall, J.(2008)Using mobile technologies for assessment and learning in practice settings: A case study.Nurse Education in Practice 8, 197–204.
[20] Binsaleh, M., &Binsaleh, S. (2013).Mobile learning: The case study of the four southern most provinces of Thailand in transforming critical to opportunity.Procedia—Social and Behavioral Sciences 91, 322–330.
[21] Jones, A. C., Scanlon, E., &Clough, G. (2013)Mobile learning: two case studies of supporting inquiry

learning in informal and semiformal settings.Computers& Education 61, 21–32.

[22] Pimmer, C., Brysiewicz, P., Linxen, S., Walters, F., Chipps, J., &UrsGröhbiel.(2014). Informal mobile learning in nurse education and practice in remote areas—acase study from rural South Africa.Nurse Education Today.Manuscript submitted for publication.

[23] Liyana Shuib, Shahaboddin Shamshirband, Mohammad Hafiz Ismail," A review of mobile pervasive learning: Applications and issues", Computers in Human Behavior 46 (2015) 239–244.

[24] Nazatul Aini Abd Majid, Hazura Mohammed and Rossilawati Sulaiman, "Students' perception of mobile augmented reality applications in learning computer organization", Procedia - Social and Behavioral Sciences 176 (2015) 111 – 116.

[25] Sasitorn Lijanporn, Jintavee Khlaisang, "The development of an activity-based learning model using educational mobile application to enhance discipline of elementary school students", Procedia - Social and Behavioral Sciences 174 (2015) 1707 – 1712.

[26] Ketty Chachil, Adeline Engkamat, Adib Sarkawi, Awang Rozaimi Awang Shuib, " Interactive Multimedia-based Mobile Application for Learning Iban Language (I-MMAPS for Learning Iban Language) ", Procedia - Social and Behavioral Sciences 167 ( 2015 ) 267 – 273.

[27] Falvo, V.; Duarte Filho, N.F.; Oliveira, E.; Barbosa, E.F. "A contribution to the adoption of software product lines in the development of mobile learning applications",2014 IEEE Frontiers in Education Conference (FIE), pp. 1-8

[28] Jaschke, S., " Mobile learning applications for technical vocational and engineering education: The use of competence snippets in laboratory courses and industry 4.0", International Conference on Interactive Collaborative Learning (ICL), 2014, pp. 605-608.

[29] Vrana, Radovan, "The developments in mobile learning and its application in the higher education including libraries", 38th International Convention on Information and Communication Technology, Electronics and Microelectronics (MIPRO), 2015, pp. 881-885.

[30] Rüdel, C. (2006). A work in progress literature survey on mobile learning and podcasts in education IMPALA Project.School of Education, University of Leicester, United Kingdom.

[31] Boyes, M. (2011,December).24 benefits of mobile learning. Retrieved from http://insights.elearningnetwork.org/?p=507

[32] Ak, Ş. (2015). The role of technology-based scaffolding in problem-based online asynchronous discussion. British Journal of Educational Technology.

[33] MacKenzie, I. S., & Tanaka-Ishii, K. (2010). Text entry systems: Mobility, accessibility, universality. Morgan Kaufmann.

[34] Shen, R., Wang, M., & Pan, X. (2008). Increasing interactivity in blended classrooms through a cutting-edge mobile learning system. British Journal of Educational Technology, 39(6), 1073-1086.